\documentclass[smallextended]{svjour3}       
\smartqed  
\usepackage{graphicx}
\usepackage{amsmath}
\usepackage{amssymb}

\usepackage{xspace}

\newcommand{\dali}{\textsc{dali}\xspace}
\newcommand{\apurva}{\textsc{a\_purva}\xspace}
\newcommand{\dast}{\textsc{dast}\xspace}
\newcommand{\matras}{\textsc{mat\-ras}\xspace}
\newcommand{\ssap}{\textsc{ssap}\xspace}

\newcommand{\NP}{\textsl{NP}}
\newcommand{\eg}{\emph{e.g.}\xspace}
\newcommand{\ie}{\emph{i.e.}\xspace}

\journalname{Optim Lett}
\begin{document}

\title{Algorithm engineering for optimal alignment of protein structure distance matrices
}

\titlerunning{Optimal alignment of protein structure distance matrices}        

\author{Inken Wohlers         \and
        Rumen Andonov         \and
	Gunnar W.\ Klau
}


\institute{Inken Wohlers \and Gunnar W. Klau \at
	    CWI, Life Sciences group, Amsterdam, the Netherlands\\
            \email{\{inken.wohlers, gunnar.klau\}@cwi.nl}  \\ \\
	    Rumen Andonov \at
            INRIA Rennes - Bretagne Atlantique and University of Rennes 1, France \\
            \email{randonov@inria.fr}
}

\date{Received: date / Accepted: date}

\maketitle

\begin{abstract}
Protein structural alignment is an important problem in computational biology. In this paper, we present first successes on provably optimal pairwise alignment of protein inter-residue distance matrices, using the popular \dali scoring function. We introduce the structural alignment problem formally, which enables us to express a variety of scoring functions used in previous work as special cases in a unified framework. Further, we propose the first mathematical model for computing optimal structural alignments based on dense inter-residue distance matrices. We therefore reformulate the problem as a special graph problem and give a tight integer linear programming model. We then present algorithm engineering techniques to handle the huge integer linear programs of real-life distance matrix alignment problems. Applying these techniques, we can compute provably optimal \dali alignments for the very first time.
\end{abstract}

\section{Introduction} 

Proteins are chains of amino acid residues that fold into complex three-dimen\-sion\-al (3D) structures. These structures largely determine the biological function of a protein. Structural alignment is the problem of detecting structural similarities between two proteins---similarities that indicate, for example, a similar biological function or a common evolutionary origin. Due to numerous applications in biology and medicine, it is an extremely important problem in computational biology. Unlike the related problem of sequence alignment, for which polynomial algorithms exist, structural alignment is computationally difficult. Current structural alignment algorithms fall in one of two categories: (i) they aim at achieving a low RMSD (root mean square deviation) of the superpositioned substructures for, ideally, a large number of aligned residues, or, (ii) they are based on aligning the inter-residue distances of the two proteins. Given the distances between all residues of a protein, its 3D structure can, except for chirality, be precisely reconstructed~\cite{Havel_1983}. Inter-residue distance matrices are thus well-suited and widely used representations to align protein structures, and used in numerous programs.

From an algorithmic perspective, structural alignment algorithms can be divided into heuristics~\cite{DiLena2010,Holm1993,Kawabata2000} or exact methods~\cite{Andonov2011,Caprara2004,DAST_SEA_2010,Wohlers2010,Xie2007}. While the first are typically fast, they do not provide a theoretical quality guarantee of their results, let alone for optimality. Exact methods compute provably optimal solutions, but may suffer from performance problems. Among the exact approaches, methods that maximize the number of short aligned distances, or the \emph{contact map overlap} (CMO), have been shown to be the fastest~\cite{Andonov2011,DiLena2010}. However, the weak point of CMO approaches is that the scoring function is too basic~\cite{Wohlers2010}. The other extreme is the sophisticated \dali (\textsc{d}istance-matrix \textsc{ali}gnment) scoring function~\cite{Holm1993}. Targeting higher precision, it takes into consideration all inter-residue distances and assigns individual scores. Yet, because of the computational complexity of structural alignment and the huge size of the problem instances, it is extremely difficult to design an exact algorithm that uses this or similarly complex scoring functions. 

In this paper, we present first successes on computing provably optimal alignments for complete inter-residue distance matrices using the \dali scoring function. We decide to focus on \dali, because it is a widely used structural alignment program with a public web service and the accuracy of its alignments is higher or among those of the top methods in various benchmarks~\cite{Kolodny2005,Mayr2007}. We introduce the structural alignment problem formally in Section~\ref{sec:probl-form}, enabling us to express a variety of scoring functions used in previous work as special cases in a unified framework. In Section~\ref{sec:mathematical-model} we propose the first mathematical model for computing optimal distance matrix alignments. We therefore reformulate the problem as a special graph problem and give a tight integer linear programming (ILP) model. We present algorithm engineering techniques to solve this model to optimality for practical instance sizes (Section~\ref{sec:algor-engin}). Finally, in Section~\ref{sec:results-discussion}, we discuss our computational results that include first provably optimal \dali alignments for non-trivial structural similarities.

\section{Problem Formulation}\label{sec:probl-form}
We now formally define the problem of aligning two protein structure distance matrices. Our general definition will allow us to express many related problems that have been addressed in previous work as special cases.

A \emph{distance matrix} $D$ of a protein of length $n$ is a symmetric $n \times n$ matrix. Rows and columns correspond to individual residues of the protein, in the order they appear in the protein sequence. The entries $D_{i,j}$, $1 \leq i \leq n$, $1 \leq j \leq n$, denote the distances between residues $i$ and $j$. Many variants are possible, \eg, distances between the $C_\alpha$ or $C_\beta$ atoms, or the minimum distance between any atoms of the residues. 

An \emph{alignment} of two distance matrices $A$ and $B$ of two proteins of length $n_A$ and $n_B$ is a matching of a subset of residues from protein $A$ with a subset of residues from protein $B$ that respects the sequential order of residues in their respective chain. More precisely, it is a pair of sequences of indices $(I, J)$ with $I = (i_1, i_2, \ldots, i_{n_L})$ and $J = (j_1, j_2, \ldots, j_{n_L})$ satisfying $|I| = |J| = n_L$, $1 \leq i_1 < i_2 < \ldots < i_{n_L} \leq n_A$, and $1 \leq j_1 < j_2 < \ldots < j_{n_L} \leq n_B$, with the interpretation that residue $i_p$ is aligned to residue $j_p$ for $p = 1, \ldots, n_L$. Such an alignment $(I, J)$ induces an alignment of pairs of inter-residue distances and we define its score as
\begin{equation}
S(I, J) := \sum_{o = 1}^{n_L}\sum_{p = 1}^{n_L} s(A_{i_o, i_p}, B_{j_o, j_p}) + S_{\text{gap}}(I, J)\label{eq:probl-formulation},
\end{equation}
where $s: \mathbb{R}^{\geq 0} \times \mathbb{R}^{\geq 0} \to \mathbb{R}$ gives the scoring term of aligning individual inter-residue distances from protein $A$ with distances from protein $B$ and $S_{\text{gap}}(I, J)$ is a function penalizing the gaps of the alignment $(I, J)$. We denote by $\mathcal{A}(A, B)$ the set of all possible alignments of two distance matrices $A$ and $B$.

The problem consists now in finding the best alignment between the two matrices with respect to $S$. It is \NP-hard, following from~\cite{Lathrop1994_1}, since variable length gaps are admitted into the alignment and interactions between amino acid residues from the sequence are admitted into the scoring function.

\begin{problem}[Distance matrix alignment]
  Given two distance matrices $A$ and $B$, find an alignment $(I, J)^*$ of $A$ and $B$ with
\[
(I, J)^* = \operatorname*{arg\,max}_{(I, J) \in \mathcal{A}(A, B)}S(I, J)\enspace.
\]
\end{problem}

A large number of algorithms exists that address the distance matrix alignment problem using a variety of scoring functions, all of which can be expressed in our framework. In the following we present some of the most popular scoring schemes.

\paragraph{Contact map overlap (CMO).} Based on a distance threshold $\tau$, two residues are defined to be in contact or not. Many algorithms have been presented~\cite{Andonov2011,Caprara2004,Xie2007} that aim at finding an alignment with the maximum number of shared contacts. This corresponds to $S_{\text{gap}}(I, J)=0$ and
\[s(A_{i, j}, B_{k, l}) = \begin{cases}
1 & A_{i, j} \leq \tau \text{ and } B_{k, l} \leq \tau\\
0 & \text{otherwise,}
\end{cases}
\]
for  $1 \leq i < j \leq n_A$ and $1 \leq k < l \leq n_B$ and some $\tau \in \mathbb{R}^{\geq 0}$.

\paragraph{\dast scoring.} Motivated by the fact that maximizing shared contacts often generates alignments with large RMSD, the \dast algorithm~\cite{DAST_SEA_2010} focuses on local alignments with low RMSD\@. This corresponds to aligning cliques of distances with low mutual distance differences and can be expressed as $S_{\text{gap}}(I, J)=0$ and 
\[s(A_{i, j}, B_{k, l}) = \begin{cases}
1 & |A_{i, j} - B_{k, l}| \leq \tau \\
-\infty & \text{otherwise,}
\end{cases}
\]
for  $1 \leq i < j \leq n_A$ and $1 \leq k < l \leq n_B$ and some $\tau \in \mathbb{R}^{\geq 0}$.

\paragraph{\dali scoring.}  \dali~\cite{Holm1993} is a popular heuristic for protein structural alignment. Expressed in our model~\eqref{eq:probl-formulation} its scoring function~\cite{Hasegawa2009,Holm1993} corresponds to $S_{\text{gap}}(I, J)=0$ and 
\[
s(A_{i, j}, B_{k, l}) =  \begin{cases}
\left(0.2-\frac{|A_{i, j} - B_{k, l}|}{\frac{1}{2}(A_{i, j}+B_{k, l})}\right) e^{-\left(\frac{1}{2}(A_{i, j}+B_{k, l})/20\right)^2} &  i \neq j \text{ and } k \neq l \\
0.2 & \text{otherwise.}
\end{cases}
\]
In our experimental results we will focus on the \dali scoring scheme.

\paragraph{\matras scoring.} Analogous to sequence substitution matrices, the structural alignment algorithm \matras~\cite{Kawabata2000}, uses log-odds value matrices $M$ as structural scores. A value $M_{\lfloor d_1 \rfloor, \lfloor d_2 \rfloor}$ indicates the log-likelihood that distance $d_1$ is aligned to distance $d_2$. A positive log-likelihood means that the corresponding distances are aligned more likely than expected by chance. In our scoring model, \matras scoring can be realized by using function $S_{\text{gap}}(I, J)$ for affine gap costs  and $s(A_{i, j}, B_{k, l}) = M_{\lfloor A_{i, j} \rfloor, \lfloor B_{k, l}\rfloor}$.

\paragraph{Protein threading problem (PTP).} It is interesting to note that local PTP~\cite{Collet2010} can also be modeled in this framework with similar log-odds-based scoring functions~\cite{Marin2002} as \matras.

\paragraph{\ssap scoring.} \ssap~\cite{Taylor1989} is an early approach to structural alignment and uses a structural scoring function that can be expressed as
\[
s(A_{i, j}, B_{k, l}) = \begin{cases}
\frac{50}{(V_{i, j}^A - V_{k, l}^B)^2 + 2} &  i \neq j \text{ and } k \neq l \\
-\left(200 \min\{n_A,n_B\}\right)^{\frac{1}{2}} & \text{otherwise,}
\end{cases}
\]
where $V_{i, j}^A$ and $V_{k, l}^B$ are vectors between residue $i$ and $j$ in protein $A$ and  $k$ and $l$ in protein $B$, resp. Using the difference of vectors for scoring,  \ssap can account for the directionality of the compared inter-residue distances. Function $S_{\text{gap}}(I, J)$ assigns a penalty of $-5$ for each gap, independent of the gap's length.

\section{Mathematical Model}\label{sec:mathematical-model}

\begin{figure}[b]
\centerline{
\includegraphics[width=\linewidth]{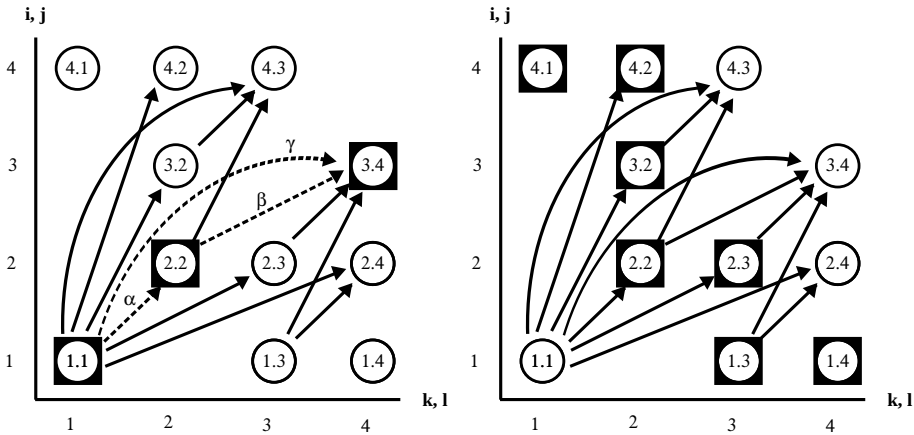}}
\caption{Alignment graph for nodes that remain after preprocessing. Left: the set of nodes $\{1.1, 2.2, 3.4\}$ forms a strictly increasing path that induces a structural alignment with score $\alpha+\beta+\gamma$ plus the score of the nodes. Here, $\alpha=2s(A_{1,2},B_{1,2})$, $\beta=2s(A_{2,3},B_{2,4})$, $\gamma=2s(A_{1,3},B_{1,4})$, and the node score $s(A_{11},B_{11})+s(A_{2,2},B_{2,2})+s(A_{3,3},B_{4,4})$. Right: Nodes on the decreasing path $\{4.1, 4.2, 3.2, 2.2, 2.3, 1.3, 1.4\}$ mutually contradict.} \label{alignment_graph}%
\end{figure}

We now present a mathematical model for structural alignment of protein distance matrices. To our knowledge, it is the first model for optimal protein structural alignment in a general setting as defined above. First, we reformulate the alignment problem in a graph-theoretical framework: An \emph{alignment graph} $G=(V,E)$ of two proteins of size $n_A$ and $n_B$ is a $n_A \times n_B$ grid graph in which each node $i.k$ represents a possible alignment of a residue $i$ in protein $A$ to a residue $k$ in protein $B$, see also Figure~\ref{alignment_graph}. Each row of $G$, from bottom to top, corresponds to a residue in protein $A$, and each column, from left to right, corresponds to a residue in protein $B$. Edges in $G$ represent possible alignment of pairs of distances. An edge $(i.k,j.l)$ denotes the possible alignment of the distance between residues $i$ and $j$ in protein $A$ with the distance between $k$ and $l$ in protein $B$. Alignment graphs for structural alignment have first been used for CMO~\cite{Andonov2011}, here we adapt them to a more general setting.

We say a node $j.l$ is strictly larger than $i.k$ if and only if $j>i$ and $l>k$ and strictly smaller than $i.k$ if and only if $j<i$ and $l<k$. Edges are directed: any edge in the alignment graph has the property that its tail is strictly smaller than its head. It is thus oriented from south-west to north-east (cf. Figure~\ref{alignment_graph}). Because of the partial ordering between nodes and because there exist only edges between ordered nodes, the alignment graph is a directed acyclic graph. An alignment is a subset $\{i_1.k_1,i_2.k_2,\ldots\}$ of nodes that can be ordered such that each node is strictly larger than the previous one, \ie, $i_1 < i_2 < \ldots$ and $k_1 < k_2 < \ldots$; we then say the nodes lie on a strictly increasing path. There is a one-to-one correspondence between alignments and strictly increasing paths. Every structural alignment can be represented by nodes on a strictly increasing path together with all induced edges (\ie the associated augmented path). Its score is the sum of edge and node weights (cf. Figure~\ref{alignment_graph}, left). Determining the structural alignment of maximum score is equivalent to the maximum weight augmented path problem in the corresponding alignment graph.

Nodes in the alignment graph contradict if they are not on a strictly increasing path. A decreasing path contains a set of \textit{mutually} contradicting nodes. That is, it is a set $\{i_1.k_1,i_2.k_2,\ldots\}$ of nodes with $i_1 \geq i_2 \geq \ldots$ and $k_1 \leq k_2 \leq \ldots$ (cf. Figure~\ref{alignment_graph}, right). We denote the set of all decreasing paths by $\mathcal{C}$. Strictly smaller and strictly larger nodes exist only in strictly increasing paths and not in decreasing paths.

We introduce some further notation describing the neighborhoods of a node $i.k$. By $V^-(i.k)$ we denote the set of nodes that are strictly smaller than $i.k$ (left neighborhood), and by $V^+(i.k)$ the set of nodes strictly larger than $i.k$ (right neighborhood). The set $\mathcal{C}_{i.k}^-$ contains all decreasing paths in $V^-(i.k)$ and the set $\mathcal{C}_{i.k}^+$ all decreasing paths in $V^+(i.k)$.

Our integer linear programming formulation uses two types of variables. Binary variables $x_{ik}$ represent nodes in the alignment graph and indicate whether residues $i$ and $k$ are aligned. Variables $y_{ikjl}$ denote whether the alignment graph edge $(i.k,j.l)$ is present in the solution, \ie whether distance $A_{i,j}$ is aligned with distance $B_{k,l}$.
\begin{align}
\hspace*{-.1em}\max\:&\sum_{i=1}^{n_A-1}\sum_{j=i+1}^{n_A}\sum_{k=1}^{n_B-1}\sum_{l=k+1}^{n_B}2s(A_{i,j}B_{k,l})y_{ikjl}+\sum_{i=1}^{n_A}\sum_{k=1}^{n_B}s(A_{i,i}B_{k,k})x_{ik}\label{eq:1}
\hspace*{-19em}\\
\hspace*{-.1em}\text{s.t.\: }&x_{ik}\geq \sum_{j.l \in C}y_{ikjl} \quad \forall C \in \mathcal{C}_{i.k}^+, i \in[1,n_A-1], k\in[1, n_B-1] \label{clique_c1}\\
\hspace*{-.1em}&x_{ik}\geq \sum_{j.l \in C}y_{jlik} \quad \forall C \in \mathcal{C}_{i.k}^-, i \in[2,n_A], k\in[2, n_B] \label{clique_c2}\\
\hspace*{-.1em}&x_{ik}\leq 1+\hspace*{-2em}\sum_{\substack{j.l \in C \\s(A_{i, j}, B_{k, l}) \leq 0}} \hspace*{-1.8em} (y_{ikjl}-x_{jl}) \quad \forall C \in \mathcal{C}_{i.k}^+, i \in[1,n_A-1], k\in[1, n_B-1]\label{clique_c3}\\
\hspace*{-.1em}&\sum_{i.k \in C} x_{ik}\leq 1 \quad \forall C \in \mathcal{C} \label{clique_c4}\\
\hspace*{-.1em}&\mathbf{x} \mbox{\ binary}, \quad \mathbf{y} \geq 0\label{cuttingplane_model}
\end{align}

Note that the objective function~\eqref{eq:1} of this ILP only models the structural part of~\eqref{eq:probl-formulation}, and not $S_{\text{gap}}(I, J)$. It is possible to integrate also linear or affine gap costs. For the sake of simplicity, we do not describe this here but refer to~\cite{Althaus2002} where this is done for multiple sequence alignment.    

Constraints~\eqref{clique_c4} guarantee that the $x$ variables form a proper alignment, \ie, a strictly increasing path in $G$. Constraints~\eqref{clique_c1} and~\eqref{clique_c2} link $x$ and $y$ variables. They prevent activating edges for which source or target node are not activated as well. Similarly, inequalities~\eqref{clique_c3} force the activation of edges whose endpoints are activated. This is a novel class of constraints and necessary because edges $(i.k, j.l)$ with $s(A_{i, j}, B_{k, l}) < 0$ would otherwise never be part of an optimal  solution. Note that all inequality classes have exponential size. Every feasible solution of model~(\ref{eq:1})-(\ref{cuttingplane_model}) is a structural alignment and constitutes in the alignment graph a strictly increasing path with its induced edges.

\section{Algorithm Engineering}\label{sec:algor-engin}

In this section, we describe several algorithm engineering techniques to solve the integer linear program from the previous section to optimality for practically relevant instance sizes. We tried out several approaches, including solving a less tight but polynomially-sized model, developing a branch-and-cut approach for the full model, combinations of the two models, several preprocessing techniques, and a divide-and-conquer approach. Due to space limitations, we only describe the most successful approach here, which is a variable elimination preprocessing step followed by a branch-and-cut approach for the full model.

\subsection{Preprocessing Using Variable Elimination}

We reduce the number of alignment graph nodes by over-estimating the score of the best structural alignment including the respective node. If this overestimated score is less than the score of a known feasible solution, the corresponding node and its adjacent edges are fathomed, \ie,  deleted from the alignment graph. The efficiency of this process depends on the quality of the used lower bound. We are currently using the heuristic solution provided by \dali~\cite{Holm1993}.

In order to compute overestimations we use double dynamic programming, similar as in~\cite{Andonov2011}. In the first, local, dynamic program, we compute a profit $p_{i.k}$ for any node ${i.k} \in V$, which corresponds to the largest value that this node can add to the global score. To compute $p_{i.k}$  we focus only on incoming and outgoing edges from $i.k$,  and consider only those with score $s(A_{i, j}, B_{k, l})>0$. These scores are then assigned to the corresponding nodes $j.l \in \mathcal{V}^-(i.k)$ and $\mathcal{V}^+(i.k)$, respectively. The weight of the heaviest strictly increasing path  $p_{i.k}^+$ in $\mathcal{V}^+(i.k)$ (resp.  $p_{i.k}^-$ in $\mathcal{V}^-(i.k)$) can be computed in time proportional to the size of the rectangles $\mathcal{V}^+(i.k)$ (resp. $\mathcal{V}^-(i.k)$) using dynamic programming. Finally, we set  $p_{i.k}=p_{i.k}^-+p_{ik}^+$.

The second  level, global, dynamic program consists in overestimating any solution containing  a given node $m.n$, \ie, in which $x_{m.n}=1$. For this purpose  we associate profits $p_{i.k}$ to all nodes, and using them as weights we compute the heaviest strictly increasing path in $\mathcal{V}^-(m.n)$ (respectively  in $\mathcal{V}^+(m.n)$). Adding the weights of these two paths, as well as $p_{m.n}$, we obtain an  overestimated score for a node $m.n$.

The efficiency  of this variable elimination depends on the preciseness  of the profit computation. The coarsest but also fastest method to potentially eliminate a node $m.n$ is to use the previously  computed profits $p_{i.k}$. The advantage is that the profits can be computed once for all nodes in the beginning. However, better upper bounds can be obtained by using profits $p^{m.n}_{i.k}$ that are computed in the same way as $p_{i.k}$ but considering only nodes in $\mathcal{V}^-(m.n) \bigcup \mathcal{V}^+(m.n)$. This fine filtering gives tighter overestimations, but is more time-consuming.

\begin{figure}
  \centerline{\includegraphics[width=\linewidth]{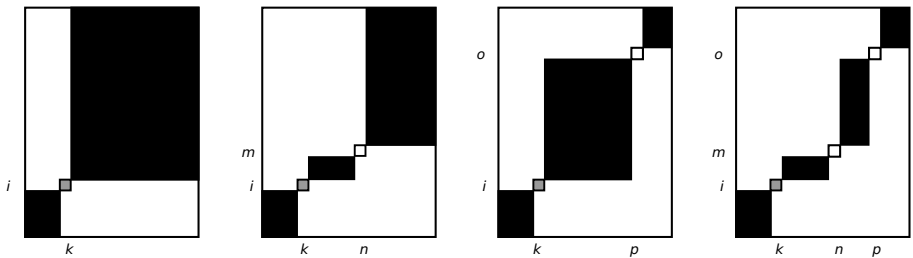}}
  \caption{Different computations for profit $p_{i.k}$ of node $i.k$. In black the feasible nodes. Left: Coarse profit computation. In black $\mathcal{V}^-(i.k)$ and $\mathcal{V}^+(i.k)$. Second from left: Fine profit computation for filtering node $m.n$ when $i.k<m.n$. $p_{i.k}^-$ stays unchanged, but $p_{i.k}^+$ can be recomputed by considering only nodes in $\mathcal{V}^+(i.k) \bigcap \left( \mathcal{V}^-(m.n) \bigcup \mathcal{V}^+(m.n) \right)$. Second from right: The same fine profit computation for filtering $o.p$. For filtering edge $(m.n,o.p)$, we can take the minimum of the two cases. Right: Profit computation for $i.k$ for fine filtering of edge $(m.n,o.p)$.} \label{fig_profits}
\end{figure}

Besides eliminating nodes, we can also eliminate edges. We implemented two ways of accomplishing this. One way, the so called coarse filtering of edge $(m.n,o.p)$, uses the same profits as for coarse  node filtering, cf. Figure~\ref{fig_profits} left. We can do better  by re-using the profits $p^{m.n}_{i.k}$ and $p^{o.p}_{i.k}$ computed during the fine filtering for nodes  $(m.n)$ and $(o.p)$.  Instead of $p_{i.k}$, we use in this case updated profits $\hat{p}_{i.k}$ defined as  $\hat{p}_{i.k}=\min (p^{m.n}_{i.k}, p^{o.p}_{i.k})$.  If $p^{m.n}_{i.k}$ and $p^{o.p}_{i.k}$ are  stored  during the fine node filtering, such a finer  edge filtering is then not more expensive than coarse filtering.

\subsection{Branch-and-Cut}

We use a cutting plane algorithm~\cite{Schrijver_1986} within a CPLEX branch-and-cut framework to solve ILP~\eqref{eq:1}--\eqref{cuttingplane_model}. We first solve the LP relaxation of an initial ILP (with reduced number of constraints~\eqref{clique_c1}--\eqref{cuttingplane_model}). In case of fractional solution, we solve a separation problem that yields a violated inequality cutting off this solution. In practice, often several of such cutting planes are generated. The new constraints are added to the relaxed ILP problem which is solved again, and so on. If no more cutting planes can be found, the formulation comprising the new constraints is solved by CPLEX within a branch-and-bound algorithm.

We now show how cutting planes for all types of constraints in model \eqref{eq:1}--(\ref{cuttingplane_model}) can be generated using Dijkstra's (or general shortest path)  algorithm in a directed acyclic graph. For inequalities~(\ref{clique_c4}) this is identical to~\cite{Lenhof1998}. For the remaining inequalities the principle is the same.

Cutting planes for inequalities~(\ref{clique_c4}) can be generated efficiently by assigning value $x_{ik}$ obtained by the linear relaxation as weight to each alignment graph node. We can determine the most violated constraint by computing the maximum weight decreasing path $C$ in the alignment graph nodes $V$. If this weight is greater than one, the nodes on the decreasing path violate constraint~(\ref{clique_c4}). Since node weights are between zero and one, we can always lift a constraint of type~(\ref{clique_c4}) by adding $x$-variables that refer to a corresponding decreasing path of maximum cardinality. Hence, we search for the maximum weight path amongst all decreasing paths of length  $(n_A \times n_B)-1$. We reformulate this problem to a minimum weight path problem by assigning weights $(1-x_{ik})$ to the nodes. Since all weights are positive, we can use Dijkstra's algorithm to solve this problem.  If for the path  $C$ found in this way  $\sum_{i.k \in C} x_{ik} > 1$ holds, it generates a cutting plane.

Cutting planes of type~(\ref{clique_c1}) and~(\ref{clique_c2}) are generated analogous to cutting planes of type~(\ref{clique_c4}). For each node $i.k$ we generate up to two cutting planes, one for incoming edges and one for outgoing edges. In the first case, we assign for fixed $i.k$ the weights $y_{jlik}$ to each node $j.l \in \mathcal{V}^-(i.k)$, and in the second case we assign $y_{ikjl}$ to each node $j.l \in \mathcal{V}^+(i.k)$. Then we compute the maximum weight decreasing path in $\mathcal{V}^-(i.k)$ and $\mathcal{V}^+(i.k)$, respectively. If this path has weight greater than $x_{ik}$, we identified a cutting plane. The reformulation of the maximum weight path problem to a minimum weight path problem that can be solved with Dijkstra's algorithm is analogous to cutting plane generation for constraints~(\ref{clique_c4}). In worst case we identify $2(n_A-1)(n_B-1)$ violated inequalities~(\ref{clique_c1}) and~(\ref{clique_c2}) in each iteration.

Similarly, we identify violated activation constraints~(\ref{clique_c3}). For each node $i.k$, we assign the weight $y_{ikjl}-x_{jl}$ to each node $j.l \in \mathcal{V}^+(i.k)$ with objective function coefficient $s(A_{i, j}, B_{k, l}) \leq 0$. Note that, because we consider only edges with negative objective function score, in this case we do not know the cardinality of the path beforehand. If the weight of the lightest decreasing path $+1$ is smaller than $x_{ik}$, we identified a violated inequality~(\ref{clique_c3}). Since we compute the minimum weight path in a directed acyclic graph with edge weights less than or equal to zero, we can not apply Dijkstra's algorithm. Instead we traverse all nodes in topological order, which is provided by sorting according to the above defined order on the nodes. A constraint of type~(\ref{clique_c3}) only cuts off the current solution if its $x_{ik}$ value is greater than zero. In practice, a good strategy is to generate cutting planes~(\ref{clique_c3}) only for nodes $i.k$ with $x_{ik}=1$, because the alignment is mainly determined by edges $(i.k,j.l)$ with positive score $s(A_{i, j}, B_{k, l})$ and there are comparatively very few edges with negative score in a solution.

\section{Results and Discussion}\label{sec:results-discussion}

\subsection{Data Sets and Setup of Computational Study}

We evaluate variable elimination and the branch-and-cut approach on the Skolnick data set, a benchmark for clustering proteins. It consists of 40 protein chains of length between 97 and 255 residues that belong to five different SCOP (structural classification of proteins) families. This data set has been extensively used for the evaluation of CMO algorithms, \eg, in~\cite{Andonov2011,Caprara2004,Jain2007,Pelta2008}. \dali skips hetero atom records and residues with incomplete backbone. Therefore we edited 7 PDB (protein data bank) files of the Skolnick data set; we changed heteroatom records to atom records and excluded residues that were for other reasons ignored during \dali computation. Furthermore, in order to be in line with \dali, we consider only the first decimal place of C$_\alpha$ atom coordinates. The remaining computations were carried out with double precision leading to slight differences between the overall score reported by \dali and the recomputed score.

To cluster the proteins in the Skolnick benchmark, we compute for each pair of proteins an alignment and a corresponding similarity score, resulting in 780 alignments. We compute them using \dali and analyze the results. \dali evaluates similarities using an empirical $z$-score which is based on the alignment's \dali score. A $z$-score above 8 yields good structural superpositions~\cite{Holm2006}. In the Skolnick data set there are 164 alignments with $z$-score greater than 8. They all correspond to pairs of proteins from the same family and are considered ``easy'' instances in~\cite{Andonov2011}. All alignments between pairs of proteins from different families have a $z$-score of less than 4 (or \dali detects no similarity). This gap between alignments with high $z$-score and alignments with low $z$-score is promoting the well-known fact that the Skolnick data set is a rather easy benchmark~\cite{DiLena2010}. We focus in the subsequent analysis only on alignments with $z$-score greater than 8 because the poor performance of variable elimination for dissimilar proteins renders it currently infeasible to fit the problems into memory.

Furthermore, we use alignments from Sisy~\cite{Andreeva2007,Berbalk2009}, a more challenging data set that has been designed with the objective to provide difficult structural alignment instances. These alignments are individually inspected by experts and essentially manually created; therefore we consider them gold-standard reference alignments.

In order to compute optimal \dali alignments using our branch-and-cut approach, we use CPLEX version 12.1 and a maximum running time of 30 hours. Instead of generating cutting planes for inequalities~(\ref{clique_c4}), we use a polynomial number of constraints as initial ILP\@,

\begin{equation}
\sum_{l=1}^k x_{il}+\sum_{j=1}^{i-1}x_{jk} \leq 1, \quad i \in[1,n_A], k\in[1, n_B]\enspace.
\end{equation}

These constraints from~\cite{Andonov2011} describe the set of increasing paths. We then add cutting planes~(\ref{clique_c1}), (\ref{clique_c2}), and~(\ref{clique_c3}) within a CPLEX cutting plane callback function. Alignments have been computed on cluster nodes each equipped with two quad core 2.26 GHz Intel Xeon processors and 24 GB of main memory running 64 bit Linux.

\subsection{Importance of Preprocessing}

Real-life problem instances for aligning complete inter-residue distance matrices are huge, consisting of $n_An_B$ $x$-variables and ${n_A \choose 2}{n_B\choose 2}$ $y$-variables. We therefore have to rely on a very effective preprocessing. In this work, we present variable elimination as preprocessing in order to minimize the number of variables that have to be considered explicitly in the \dali ILP model. The effectiveness of this preprocessing depends on the similarity of the two proteins; if we apply preprocessing to two identical proteins, only the $x$-variables denoting identical residues and the $y$-variables of corresponding pairs of distances remain. As expected, the percentage of $x$-variables eliminated during preprocessing thus correlates with the \dali $z$-score of the alignment (correlation coefficient $0.91$, data not shown). The percentage of eliminated $x$-variables can be used successfully for correct classification of the Skolnick data set. This is an indication that highly similar proteins are easy to align optimally---an observation first made with exact CMO alignment algorithms.

Variable elimination has to be effective, firstly in order to reduce the model such that it fits into memory and secondly such that we do not have to add too many cutting planes. This is oftentimes feasible for the class of alignments that leads to good structural superposition ($z$-score greater than $8$). For pairs of proteins with more subtle structural similarities, for example if only subsets of both proteins are structurally similar, this preprocessing fails. For these cases, the current variable elimination is a good first step, but it is not sufficient to reduce the model such that it can be handled by CPLEX.

\subsection{Optimal Distance Matrix Alignments}

Our branch-and-cut approach solves 75 of the 164 Skolnick alignments (46\%) to provable optimality. The running times for the solved instances vary between 24s and 70296s and depend on the number of $y$-variables after variable elimination. The biggest solved instance has little less than 3 million $y$-variables. For 23 protein pairs (14\%), CPLEX runs out of memory before the time limit is reached. Those instances have more than 15 million pairs of distances. 

For 32 of the solved Skolnick instances (43\%), the heuristic solution provided by \dali was proven to be optimal. In the remaining instances, the optimal solution improved the heuristic solution slightly (less than 2\% improvement in \dali score). With improvements in this order of magnitude, \dali might not fail to produce the optimal alignment, but determine a different alignment because it possibly uses different precision during computations. Furthermore, the compared Skolnick proteins are very similar in length and structure and the optimal alignment usually globally aligns them with a few small gaps. Therefore a good performance of the \dali program is expected. We assume that in more difficult instances, in which the optimal alignment is not global, there is more room to improve on the heuristic \dali solution. 

We use small Sisy alignments in order to evaluate our branch-and-cut approach on such more difficult instances. Four instances are solved. We prove the optimality of the heuristic \dali solution for two of them and for two improve with our exact solution the heuristic \dali score slightly (by less than $1\%$).

For global alignments of structurally very similar proteins, \eg, Skolnick instances, the shift from a simple scoring function like CMO, to a more sophisticated scoring of all distances using for example the \dali scoring function, shows little effect. Nonetheless, our previous evaluation of the complete Sisy data set illustrates that for detection of more complex structural similarities, basic scores as CMO are not always sufficient to obtain gold-standard reference alignments, and the \dali scoring function performs on average better~\cite{Wohlers2010}. 

\begin{figure}[h]
  \centerline{
    \includegraphics[width=0.3\linewidth]{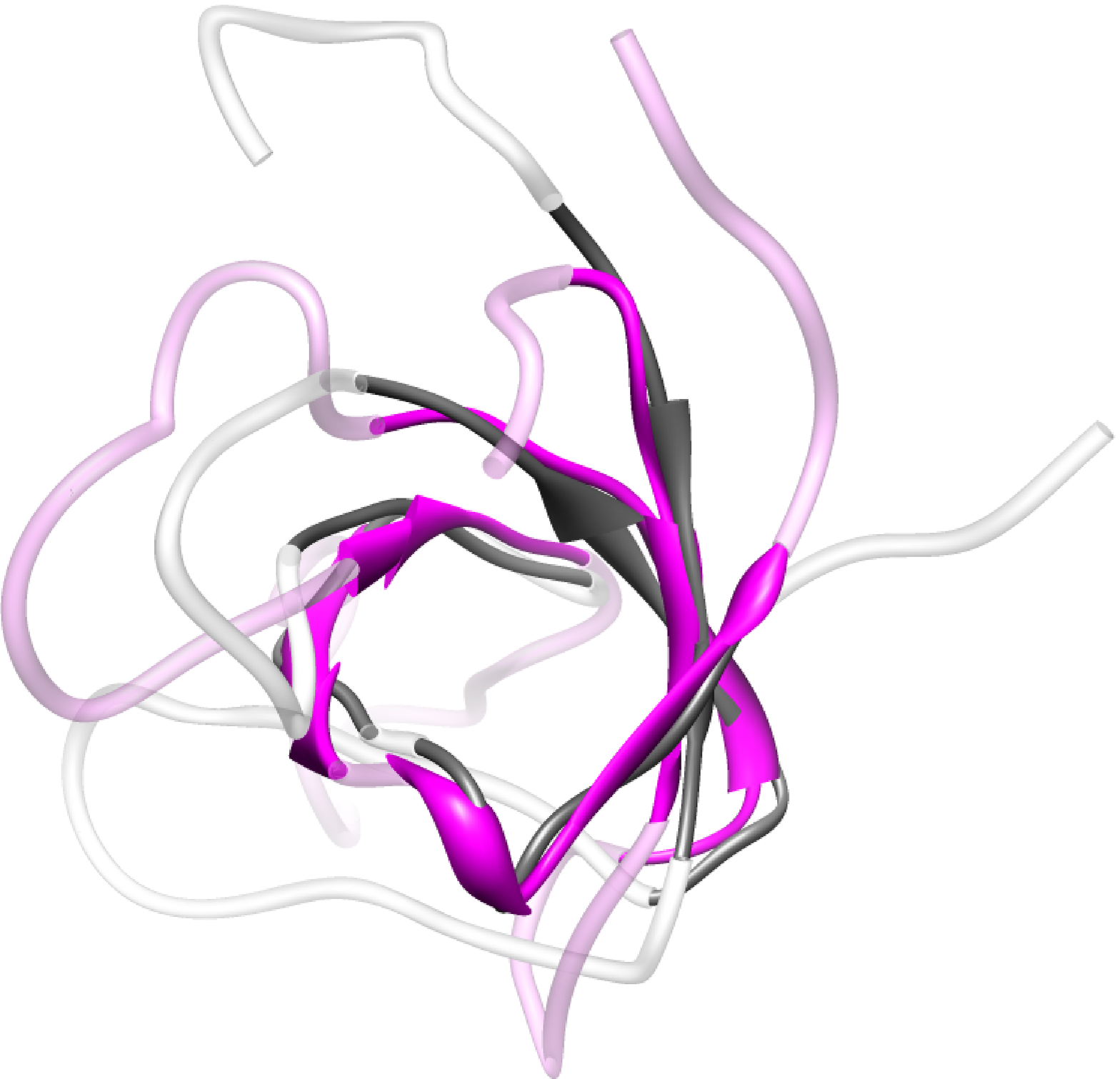}
    \hspace{0.2cm}
    \includegraphics[width=0.3\linewidth]{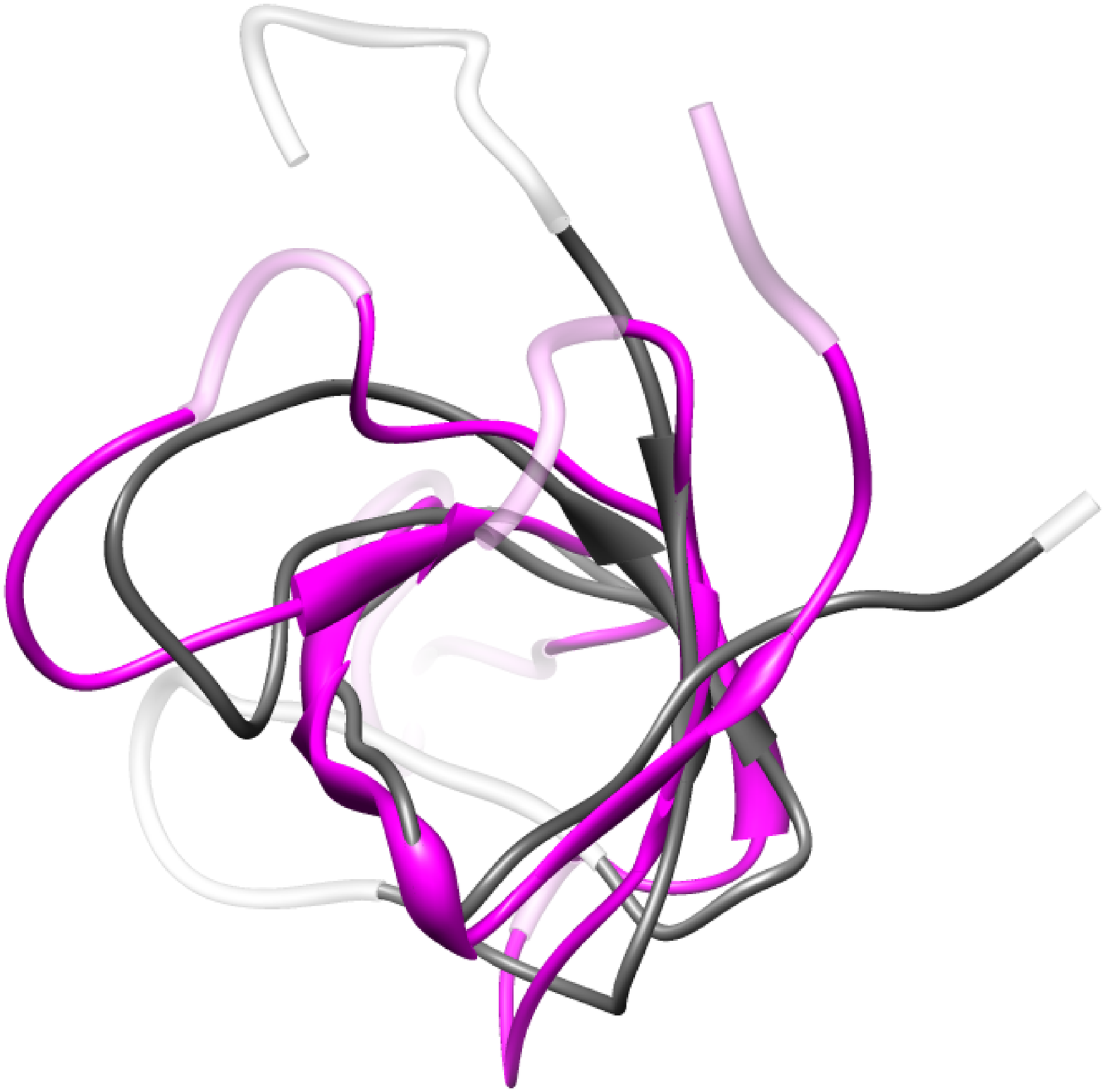}
    \hspace{0.2cm}
    \includegraphics[width=0.3\linewidth]{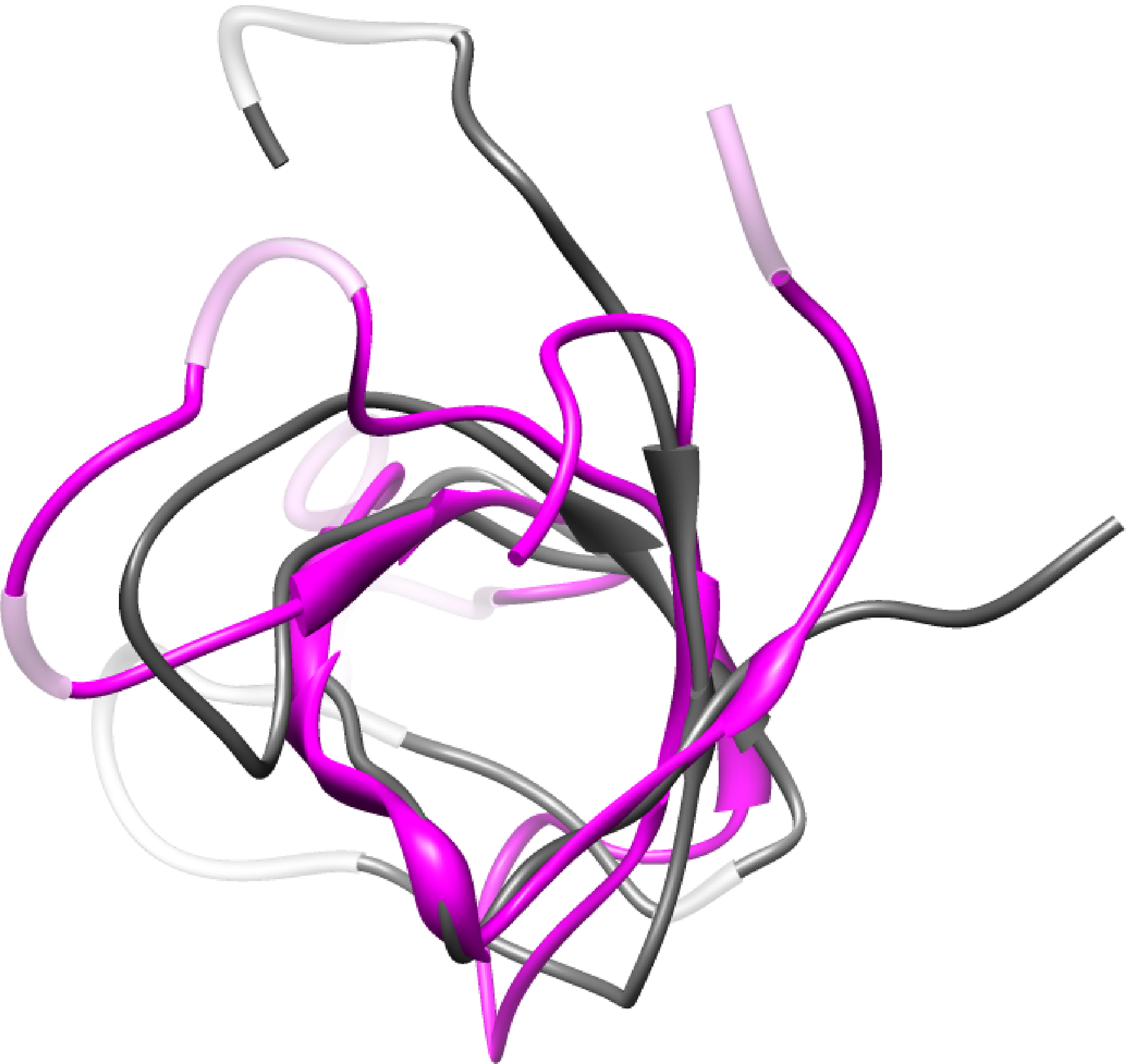}}
  \caption{Alignment of 1aawA (gray) and 1gxiE (pink), an instance of the Sisy set. Optimal superposition according to the respective alignment. Residues colored in dark tone are aligned, residues colored in light tone are unaligned. Left: The Sisy reference alignment (29 aligned residues, RMSD of 1.14). Middle: The heuristic \dali alignment correctly aligns all residues of the reference alignment, but extends the alignment length to 50 (RMSD of 2.55). Right: The optimal CMO alignment; it correctly aligns 96.55\% of the aligned residues of the reference alignment. Alignment length is 56, RMSD 4.25. Additional gaps are inserted. Overaligning and insertion of additional gaps leads to a low RMSD value.} \label{superpositions}%
\end{figure}

\subsection{Conclusion and Future Directions}

The branch-and-cut approach presented in this paper is the first exact algorithm that offers the feature of penalizing the alignment of non-compatible distances, \eg, a small and a large distance. Furthermore, it illustrates for the first time that it is feasible to compute provably optimal alignments of complete inter-residue distance matrices by means of preprocessing followed by a dedicated branch-and-cut approach that is implemented within a general purpose ILP solver. Especially, we were able to compute \dali alignments to optimality, demonstrating that this popular and widely used heuristic structure comparison method generates optimal or close-to-optimal alignments, at least for relatively similar and relatively small problem instances. 

The techniques we apply are currently successful only in the cases of similar proteins. Based on our experience, we believe that it is not realistic to provide exact solutions for all instances. On the other hand, all our results show that it is not necessary to consider all pairs of inter-residue distances in order to obtain good alignments. Methods from molecular distance geometry~\cite{Wu_2008}, for example, can uniquely reconstruct a protein's 3D structure from a small subset of distances. Current alignments of sparse inter-residue distance matrices might thus perform promising because they capture to large extend the protein's structure. Nonetheless, we observe that it is essential to penalize non-compatible distances by assigning them a negative score. An illustration is given in Figure~\ref{superpositions}: the CMO scoring function greedily aligns as many residues as possible, which leads to an increase of alignment length at the expense of precise structural similarity of compact substructures as measured by RMSD. 

Because of these observations, we plan to move towards structural alignment scoring schemes that reduce the size of the model by excluding pairs of distances. A first successful, yet still non-negative, scoring scheme was presented in~\cite{Wohlers2010}. Secondly, \apurva, the implementation from~\cite{Andonov2011}, is able to provide optimal CMO alignments even for very large and dissimilar instances. Based on these observations we will come up with hypotheses which pairs of distances can be excluded without losing alignment precision. Computing alignments to optimality, we can then objectively test these hypotheses. 

\begin{acknowledgements}
This work has been supported by DFG grant KL 1390/2--1 and was partly done when I.\ Wohlers was visiting IRISA supported by an INRIA grant. R.\ Andonov  is supported by BioWIC ANR-08-SEGI-005 project and partially by DVU-01/197/16.12.2008 NSF-Bulgaria. Computational experiments were sponsored by the NCF for the use of supercomputer facilities, with financial support from NWO.
\end{acknowledgements}

\bibliographystyle{spmpsci}
\bibliography{biblio}

\end{document}